\documentclass[12pt]{article}

\usepackage[a4paper,margin=1in]{geometry}
\usepackage{times}
\usepackage{microtype}
\usepackage{amsmath,amssymb,mathtools}
\usepackage{siunitx}
\usepackage{booktabs}
\usepackage{graphicx}
\usepackage{hyperref}
\usepackage[ruled,vlined]{algorithm2e}
\usepackage{xcolor}
\usepackage{caption,subcaption}
\usepackage{multirow}

\title{From Light to Energy: Machine Learning Algorithms for Position and Energy Deposition Estimation in Scintillator–SiPM detectors}
\author{
Yoav Simhony$^{1,2}$\thanks{yoavsimhony@mail.tau.ac.il} \and
Alex Segal$^{3}$\thanks{alexs@afeka.ac.il} \and
Ofer Amrani$^{2}$\thanks{ofera@tau.ac.il} \and
Erez Etzion$^{1}$\thanks{ereze@tau.ac.il}
}

\date{
\small
$^{1}$Tel Aviv University, Raymond and Beverly Sackler School of Physics and Astronomy\\
$^{2}$Tel Aviv University, School of Electrical Engineering\\
$^{3}$Afeka College of Engineering, Unit of Mathematics\\
}

\begin{document}
\maketitle

\abstract{Scintillator-SiPM Particle Detectors (SSPDs) are compact, low-power devices with applications including particle physics, underground tomography, cosmic-ray studies, and space instrumentation. They are based on a prism-shaped scintillator with corner-mounted SiPMs. Previous work has demonstrated that analytic algorithms based on a physical model of light propagation can reconstruct particle impinging positions and tracks and estimate deposited energy and Linear Energy Transfer (LET) with moderate accuracy. In this study, we enhance this approach by applying machine learning (ML) methods, specifically gradient boosting techniques, to improve the accuracy of spatial location and energy deposition estimation. Using the GEANT4 simulation toolkit, we simulated cosmic muons and energetic oxygen ions traversing an SSPD, we trained XGBoost and LightGBM models to predict particle impinging positions and deposited energy. Both algorithms outperformed the analytic baseline. We further investigated hybrid strategies, including hybrid boosting and probing. While hybrid boosting provided no significant improvement, probing yielded measurable gains in both position and LET estimation. These results suggest that ML-driven reconstruction provides a powerful enhancement to SSPD performance.}

\section{Introduction}
The ability to reconstruct charged particle trajectories and energy deposition with compact detectors has significant implications for cosmic-ray studies, spaceborne instrumentation, and applied particle physics. Scintillator detectors instrumented with Silicon Photomultipliers (SiPMs) are particularly attractive due to their compactness, low power consumption, and robustness against environmental conditions.
A recent contribution by \cite{Simhony2024} described the Scintillator–SiPM Particle Detector (SSPD). This detector was introduced in \cite{simhony2021tausat} and later validated through in-orbit tests aboard the International Space Station \cite{COTS-Capsule-System}. The SSPD consists of a truncated prism scintillator read out by four SiPMs at its corners. The analytic model for this configuration translated sensor intensity signals into estimates of particle impinging position and energy deposition. By applying an SSPD array, one can achieve particle track reconstruction and Linear Energy Transfer (LET) estimation, resulting in millimeter-scale localization and LET uncertainties of $5-10\%$ under favorable conditions. 

Due to the geometrical distribution of the SiPMs and due to some numerical instability of the algorithm described in \cite{Simhony2024}, it underperforms in certain areas of the scintillator, mostly for particle events localized at the extremity of the SSPD.

In parallel, the broader field has increasingly recognized the promise of machine learning methods for scintillator-based event reconstruction. For example, a 2021 study \cite{Heredge2021} demonstrated that boosted decision trees and neural networks can outperform analytic techniques in predicting event parameters from scintillation light distributions, particularly in complex or high-noise scenarios. This work demonstrated ML implementations to detector response modeling showing that data-driven methods are capable of capturing effects beyond the reach of analytic formulations. In their work, the authors compared boosting algorithms such as XGBoost, probabilistic neural networks and an analytic model in various geometries and sizes of scintillators in the case of muon particles on a track perpendicular to the scintillator's surface. They showed that the best results are achieved using a square scintillator and XGboost. 

Related research on the application of machine learning to particle tracking and detector optimization has demonstrated promising results across various detector types. For example, Yaary et al. \cite{yaari2025trees} compared tree-based algorithms and neural networks for enhancing real-time tau-lepton selection in proton–proton collisions. Additional studies have explored similar approaches in diverse contexts, including event reconstruction, and position estimation and particle classification \cite{thomadakis2022using, siviero2021first, daniel2024deep, adhikari2025position, ticchi2022embedded, buonanno2022gamma}.

Motivated by these developments, the present study seeks to combine the analytic physics-based algorithm described in \cite{Simhony2024} with machine learning algorithms. Specifically, we investigate whether gradient boosting algorithms - XGBoost and LightGBM - can improve track reconstruction and LET estimation in the SSPD configuration of \cite{Simhony2024}. Beyond direct application of boosting, we also explore hybrid strategies that integrate analytic-physics-based and ML-driven approaches, including hybrid boosting and probing \cite{didona2015enhancing}. By comparing analytic, pure ML, and hybrid models, we aim to assess the practical gains, in terms of estimation accuracy in SSPDs, achievable through machine learning.
We test our methods, similarly to \cite{Simhony2024}, on GEANT4~\cite{Agostinelli2003} simulated muon particles and high energy oxygen ions. 

\section{Methods}
\subsection{Detector Geometry and Simulation}
\begin{figure}[!htbp]
  \centering
     \includegraphics[width=\textwidth]{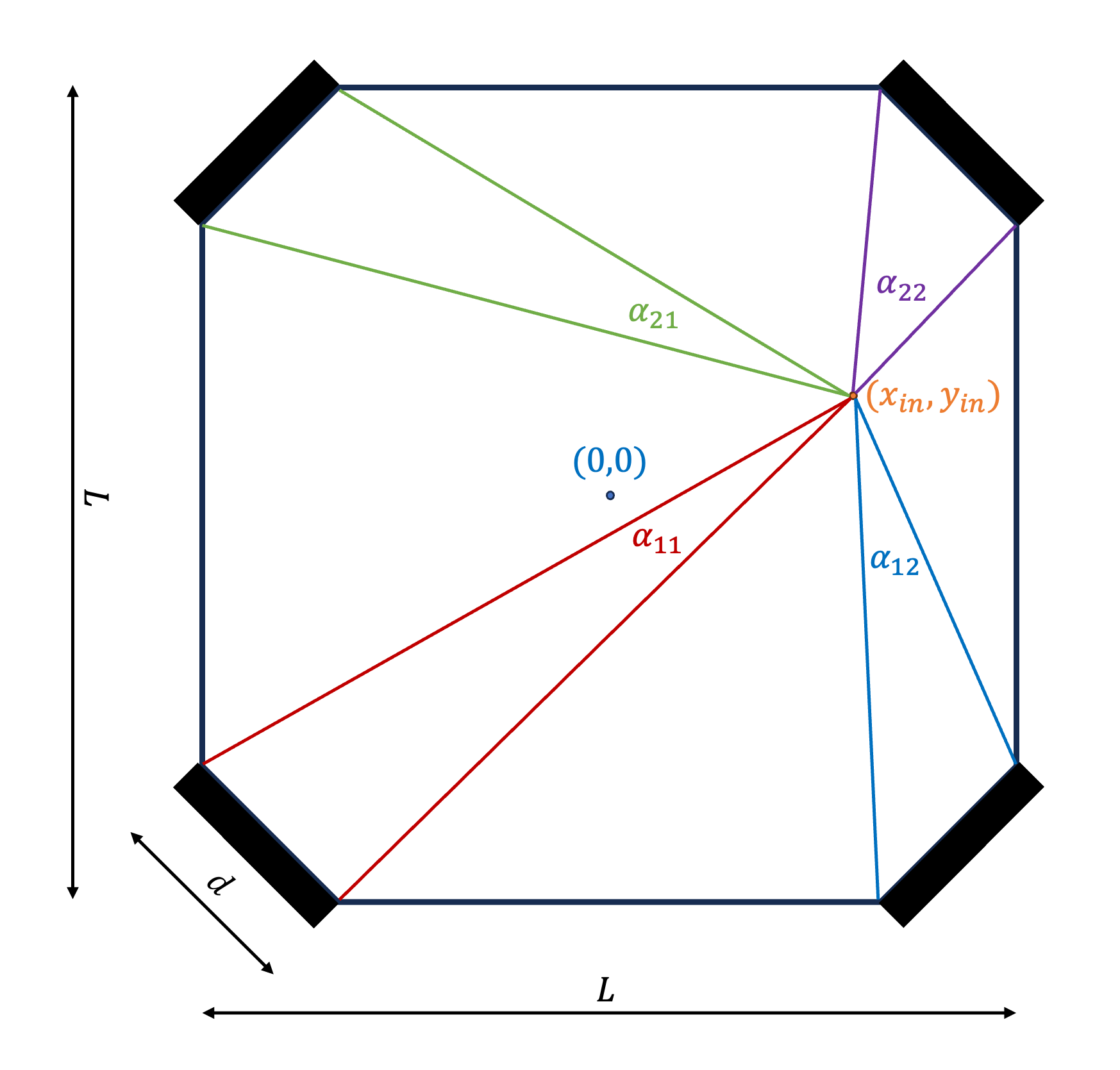}
  \caption{Top view of the SSPD \cite{Simhony2024}. The four SiPMs are positioned at the truncated corners of a prism-shaped scintillator. A particle impinging on the scintillator at position $(x_{in}, y_{in})$ is shown, along with the angles $(\alpha_{11}, \alpha_{12}, \alpha_{21}, \alpha_{22})$ defined between the impinging point and the edges of the four SiPMs.}
  \label{fig:2d_scint}
\end{figure}

We adopted the geometry from the original work in \cite{Simhony2024} (see Figure \ref{fig:2d_scint}). The detector comprises a prism-shaped polyvinyl-toluene scintillator with truncated corners (dimensions: 70~mm × 70~mm × 6.7~mm), with each corner coupled to one of four SiPMs.

The scintillator's dimensions and optical properties were modeled according to published specifications \cite{Simhony2024, eljen_scintillator}.

We generated two datasets using GEANT4 (version 10.7):
\begin{itemize}
    \item Muon dataset: $10^5$ simulated 4~GeV muons with randomized incidence positions and angles.
    \item Oxygen dataset: $10^5$ simulated 18~GeV oxygen ions with randomized incidence positions and angles.
\end{itemize}

For each simulated event, we recorded the number of photons measured by each SiPM (photons vector), the particle's energy deposition along the track within the scintillator, and the location of incidence.

\subsection{The Physics-Based Analytic model}
The physics-based analytic model (AM) introduced in \cite{Simhony2024} is based on the fact that the number of photons reaching each SiPM is proportional to the angle $\alpha_{ij}$ between the edges of the SiPM and the particle incidence location ($ \left( \alpha_{11}, \alpha_{12}, \alpha_{21}, \alpha_{22} \right) $ in Figure \ref{fig:2d_scint}):

\begin{equation}
N = C_s \cdot \textrm{LET} \cdot \frac{\alpha}{2\pi}, 
\label{eq:detector_formulas_simple}
\end{equation}
where $\textrm{LET}$ is the particle's linear energy transfer and $C_s$ is a constant dependent on the scintillator's geometry, density and efficiency in converting particle deposited energy to electromagnetic energy (photons).

Using this model, one may pre-compute the resulting measurements (up to a multiplicative constant) on a grid with a given resolution, and use the obtained results to estimate the particle's impinging location. For details see Algorithm $1$ in \cite{Simhony2024}. After the particle's impinging location is obtained, the particle's LET may be approximated using the analytic model.

The analytic approach is computationally efficient and physically interpretable, making it suitable for deployment in resource-constrained environments such as space missions. However, due to numerical instability of the algorithm, nonlinearities in events where the particle impinges the scintillator very close to a SiPM, and due to the variability of the Geometric Dilution of Precision (GDOP) across the SSPD's area it under-performs at certain areas of the scintillator, mostly near the edges, where the GDOP is small (see Figure \ref{fig:oxygen_sim_lgb}(a)).

%
\section{Machine learning methods}
%
%
\subsection{Gradient Boosted Regression (XGBoost and LightGBM)}
%
We first explore direct regression from photon vectors to incidence location using gradient boosting. Two widely used frameworks are considered:
\begin{itemize}
\item XGBoost is a highly effective tree boosting algorithm \cite{Chen2016}. It works by constructing additive ensembles of decision trees using second-order gradient information. It is robust to nonlinear feature interactions and performs well on structured datasets. Since its introduction, XGBoost, due to its scalability and speed, quickly became one of the most popular machine learning algorithms. In our context, it effectively models the nonlinear relationship between the signals detected by the SiPMs, the particle impinging positions and as a result, LET estimation.

\item Designed by Microsoft, LightGBM \cite{Ke2017} introduces histogram-based training and leaf-wise tree growth to improve training  and prediction efficiency. LightGBM is designed for faster training speeds and higher efficiency, particularly on large datasets, compared to some other gradient boosting frameworks like XGBoost. This is achieved through innovative techniques such as Gradient-based One-Side Sampling (GOSS) and Exclusive Feature Bundling (EFB), which optimize the process of finding optimal split points in decision trees.
\end{itemize}

XGBoost and LightGBM were selected because they represent state-of-the-art gradient boosting methods, well suited to capturing the nonlinear mapping between scintillation light distribution and particle impact parameters. XGBoost is robust and widely validated, while LightGBM offers improved training efficiency on large datasets. Using both models provides a robust methodological basis for evaluating gradient boosting performance in this context.
It is worth noting that both methods improve the localization accuracy compared to the analytical baseline. However, they replace rather than integrate with the physics model, potentially losing interpretability and extrapolation power in regimes outside the training distribution. 


\subsection{Hybrid intelligence: a fusion of machine learning and analytics}

To combine the advantages of the analytic model with those of machine learning algorithms, we examined two established approaches for hybrid architectures (see \cite{didona2015enhancing}).

\subsubsection{Boosting}
Boosting is a strategy that uses the analytic model as a base predictor, and corrects it iteratively by learning the residual errors. Specifically, a sequence of regressors $\Gamma_m$ is trained to predict residuals between the analytic estimate and the ground truth. These regressors, that may be of different types, are stacked in a chain, and each regressor attempts to minimize the error of the previous element in the chain. Their weighted outputs are added back to the analytic prediction:
\begin{equation}
\hat{y} =AM(X)+\sum_{m=1}^M \beta_m\cdot \Gamma_m(X),
\label{eq:hyboost}
\end{equation}
where $\beta_m$ are optimized coefficients that are trained to minimize the error up to the $m$-step. For the full algorithm, see Algorithm 2 in \cite{didona2015enhancing}.
This residual boosting approach systematically reduces bias across the detector while preserving the physical basis of the analytic model. Notice that the model is independent of the type of regressors $\Gamma_m$, which can be chosen at will. We have tested Boosting with both - XGBoost and LightGBM.

%
\subsubsection{Probing}
%
%
Probing is a hybrid strategy well-suited for scenarios where an analytic model offers fast but occasionally inaccurate predictions.
A binary classifier is trained to predict whether the analytic model’s error for a given input is below a threshold $\tau$.
At inference, if the binary classifier predicts high accuracy, the analytic model is used for prediction; otherwise, the event is routed to a machine learning regressor (e.g., LightGBM/XGboost). This selective strategy retains the speed and physics grounding of the analytic method across much of the detector while deploying machine learning only in challenging regions such as edges and corners. For full details, see Algorithm 3 in \cite{didona2015enhancing}. \\ \\
In our work, we have tested both of these approaches, with both types or regressors. Some showed a significant improvement, and some did not.

%
\section{Training and Hyperparameters}
%
%
\subsection{Inputs and Targets}
The feature vector per event is the 4-tuple measurement of photons reaching the four SiPMs (Top-Left, Bottom-Right, Bottom-Left, Top-Right). The model's target is the 2D particle impinging position within the scintillator, defined at the midpoint along the Z-axis. Prior to any ML stage (regression or classification), inputs are L2-normalized across features.
The data was split $75\%-25\%$ for training and testing, with cross-validation.
\subsection{XGBoost and LightGBM}
The XGB regressor was trained for optimizing the root mean squared error (RMSE) with $10^5$ estimators, a small learning rate of $2^{-4}$ and a maximal depth of $4$. Early stopping was not used. \\
The LightGBM regressor was trained with a max depth of $7$, 
learning rate of ~$0.01$ and roughly $1000$ estimators.
For tuning of the learning parameters we used the Optuna optimization framework (see \cite{akiba2019optuna}).

\subsection{Hybrid Boosting}
Hybrid boosting, also called iterative residue training, was implemented and tested with up to $25$ residual learners $\Gamma_m$ according to formula \ref{eq:hyboost}, and the multiplicative coefficients were optimized according to the algorithm in \cite{didona2015enhancing}.

\subsection{Probing Hybrid}
A classifier was trained that predicts whether the error of the analytic model is likely to be below a threshold $\tau$. We tested XGBoost, LightGBM and kNN classifiers. The kNN classifier with $k=2$ and Manhattan distance performed best. Using this classifier, we implemented the hybrid model that applies the analytic model if the estimated error is below $\tau$ and the machine learning method otherwise. The threshold $\tau$ was chosen for the optimal results using grid search.

\section{Results}
\subsection{Localization and LET estimation accuracy}
\begin{figure}[!htbp]
  \centering
     \includegraphics[width=\textwidth]{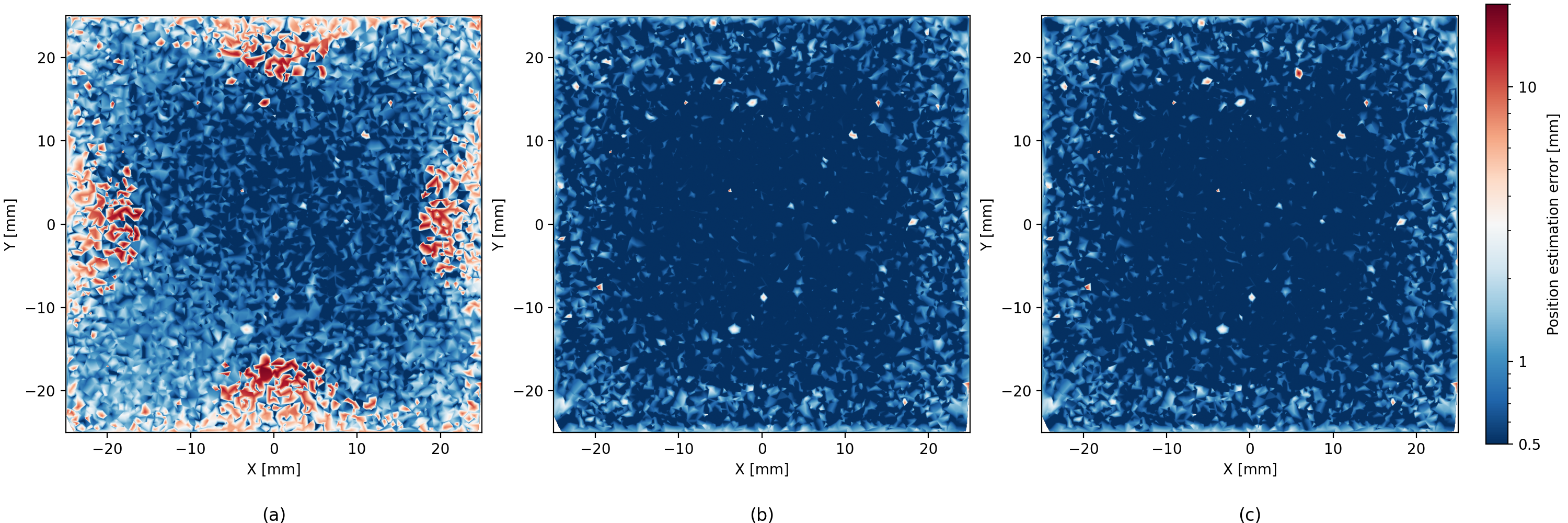}
  \caption{Comparison of localization error heatmaps in a centered $50mm \times 50mm$ area of interest of the SSPD for (a) the analytic model, (b) pure machine‑learning using LightGBM, and (c) hybrid approach with LightGBM.  Each subfigure uses the same color scale to facilitate direct comparison.}
    \label{fig:oxygen_sim_lgb_50}
\end{figure}

With the new, larger dataset, the analytic model achieved results comparable to the results in \cite{Simhony2024}, as expected. The machine learning methods, both XGBoost and LightGBM, provided a significant improvement in accuracy of roughly $30\%$ RMSE over the analytical model. The hybrid method of iterative residues learning did not improve upon the pure ML methods, however the addition of probing improved the accuracy by another $10\%$. 
The complete results are detailed in Tables \ref{tab:summary} and \ref{tab:summary2}. If we limit the particle interactions with the detector to a centered 
$50mm \times 50mm$ region of interest, the performance improves substantially, with the machine learning model yielding an average localization error of 0.52 mm, compared to 1.66 mm obtained with the analytic model (see Figure \ref{fig:oxygen_sim_lgb_50}).



\begin{table}[h]
\centering
\caption{Performance summary on \textsc{Geant4} muon simulations.}
\label{tab:summary}
\begin{tabular}{lcccc}
\toprule
 \multirow{2}{*}{Method}  & \multicolumn{2}{c}{Position RMSE [mm]} & \multicolumn{2}{c}{Mean LET error} \\
\cmidrule(lr){2-3}\cmidrule(lr){4-5}
 & Full area & 50$\times$50 & Full area & 50$\times$50 \\
\midrule
Analytic (AM)        & 6.1 & 5   & $6\%$   & $4\%$ \\
LightGBM             & 4.3 & 2   & $4.5\%$ & $1\%$ \\
XGBoost              & 4.2 & 2.05   & $4.5\%$ & $1\%$ \\
Probing LightGBM     & 4.0 & 2   & $4\%$   & $1\%$ \\
Probing XGBoost      & 4.0 & 2   & $4\%$   & $1\%$ \\
\bottomrule
\end{tabular}
\end{table}

\begin{table}[h]
\centering
\caption{Performance summary on \textsc{Geant4} oxygen simulations.}
\label{tab:summary2}
\begin{tabular}{lcccc}
\toprule
\multirow{2}{*}{Method}  & \multicolumn{2}{c}{Position RMSE [mm]} & \multicolumn{2}{c}{Mean LET error} \\
\cmidrule(lr){2-3}\cmidrule(lr){4-5}
 & Full area & 50$\times$50 & Full area & 50$\times$50 \\
\midrule
Analytic (AM)        & 3.3 & 1.6   & $4\%$   & $2\%$ \\
LightGBM             & 2.3 & 0.52   & $3\%$ & $0.3\%$ \\
XGBoost              & 2.3 & 0.53   & $3\%$ & $0.25\%$ \\
Probing LightGBM     & 1.9 & 0.51   & $2.5\%$   & $0.3\%$ \\
Probing XGBoost      & 2 & 0.51   & $2.5\%$   & $0.25\%$ \\
\bottomrule
\end{tabular}
\end{table}

%
\subsection{Error heatmaps}
%

\begin{figure}[!htbp]
  \centering
     \includegraphics[width=\textwidth]{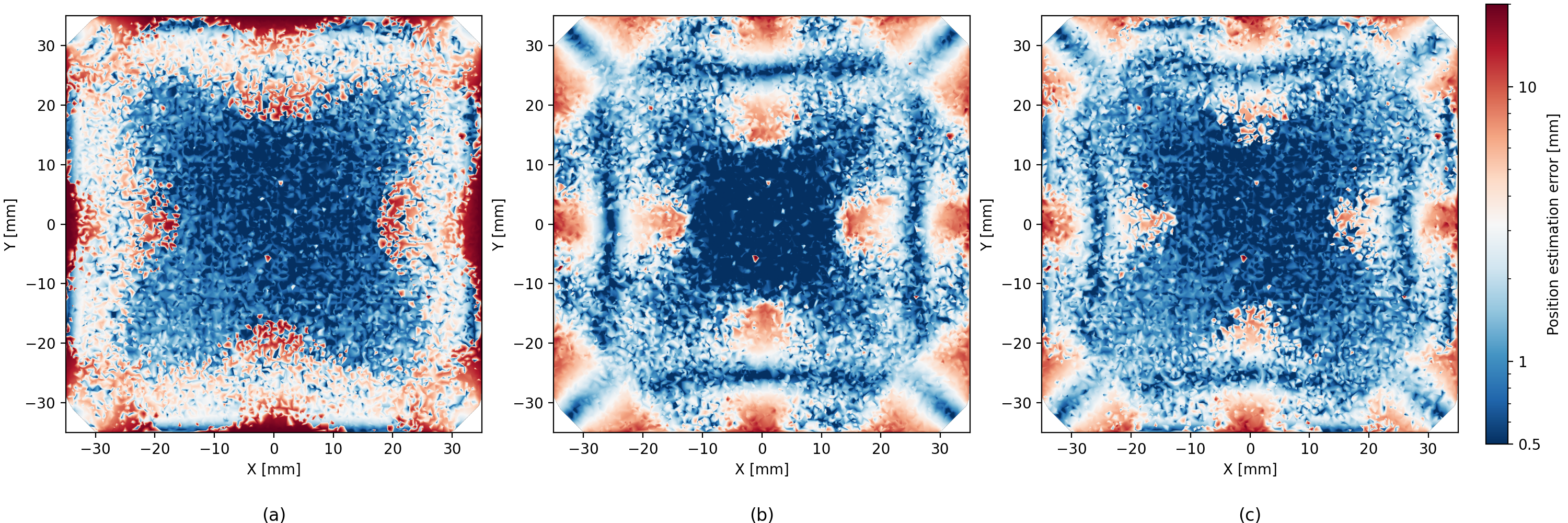}
  \caption{Comparison of localization error heatmaps for the (a) analytic model, (b) pure machine‑learning, and (c) hybrid approach.  Each subfigure uses the same color scale to facilitate direct comparison.}
    \label{fig:oxygen_sim_lgb}
\end{figure}

\begin{figure}[!htbp]
  \centering
     \includegraphics[width=\textwidth]{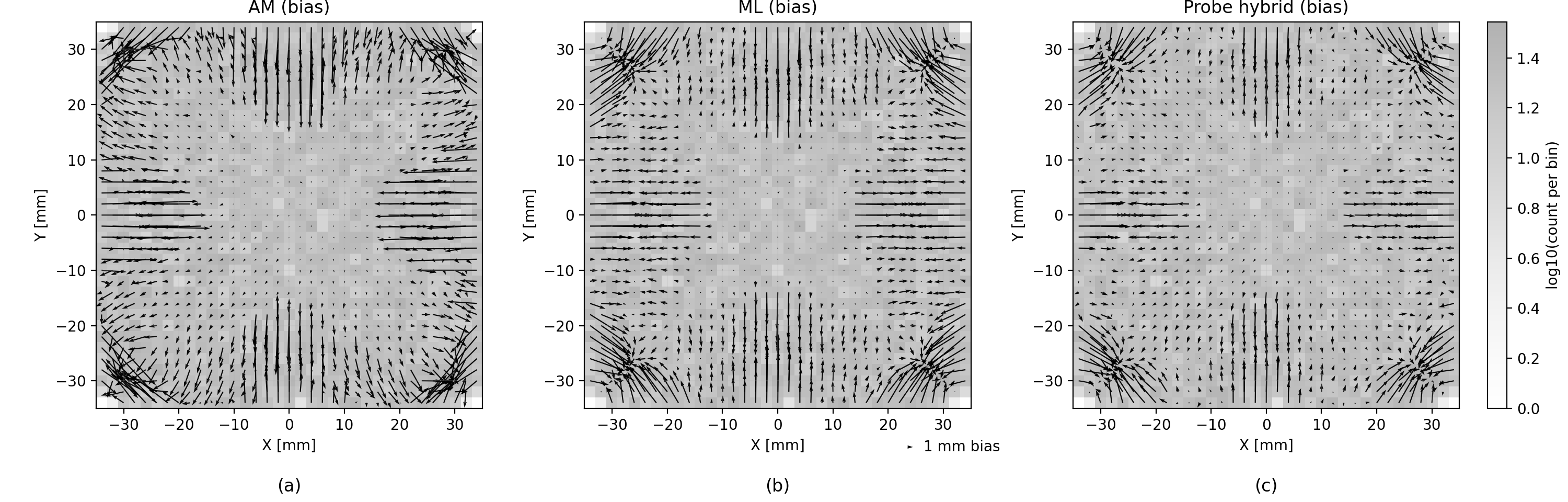}
     \caption{Mean error direction and size vector maps for the (a) analytic, (b) machine learning and (c) the hybrid models. Notice that the errors are distributed differently on each axis, depending on the location. This difference is due to variability of GDOP across the area of the scintillator and nonlinear effects very close 
     to the SiPMs.}
  \label{fig:oxygen_sim_quiver}
\end{figure}

\begin{figure}[!htbp]
  \centering
     \includegraphics[width=\textwidth]{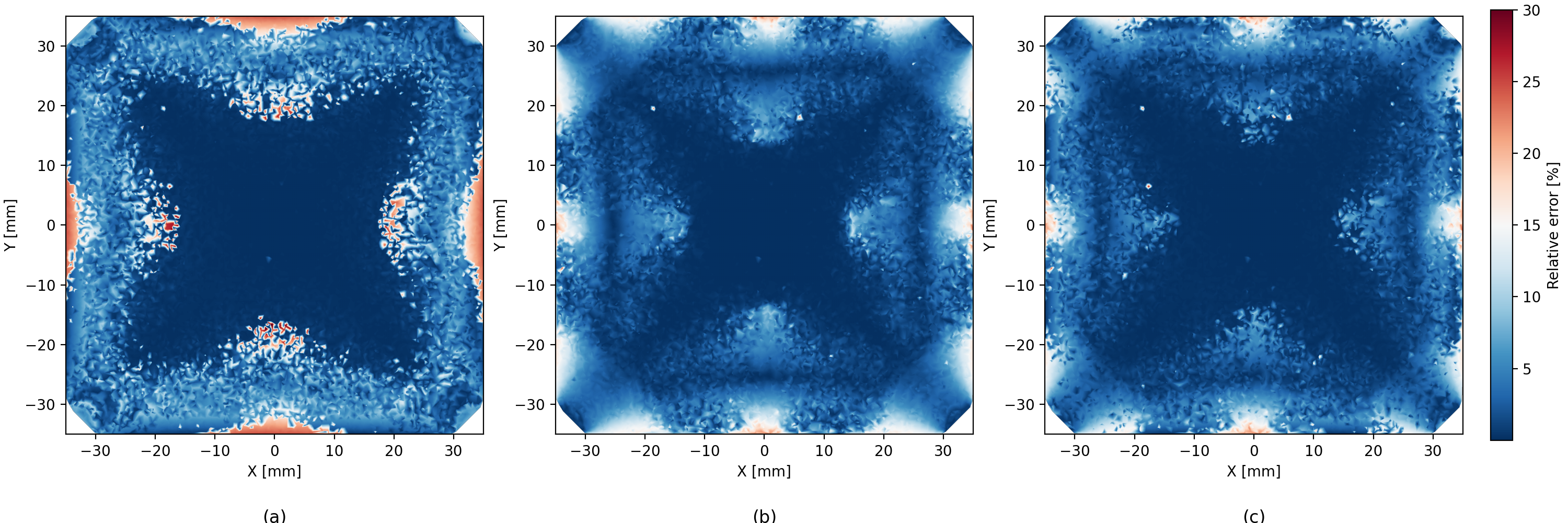}
  \caption{Comparison of relative LET error heatmaps for the analytic model (a), pure machine‑learning (b), and hybrid approach (c).  Each subfigure uses the same color scale to facilitate direct comparison.}
  \label{fig:oxygen_sim_let}
\end{figure}

Fig.~\ref{fig:oxygen_sim_lgb} shows that the precision of the analytic model 
is degraded 
near the edges of the scintillator. 
Although
machine learning methods help reduce this effect,
they tend to produce less accurate estimates
near the center area of the scintillator. 
The hybrid method seems to successfully combine both the analytic and the machine learning approaches,
and we can see an improved estimate near the center as well as 
near the edges. In addition to the error heatmap, similar effects can be seen in Figure  \ref{fig:oxygen_sim_quiver}, which shows the average error size and the direction map with different models.
A similar effect, though to a lesser degree, can be seen for LET estimates in Figure \ref{fig:oxygen_sim_let}. 

\begin{figure}[!htbp]
  \centering
     \includegraphics[width=\textwidth]{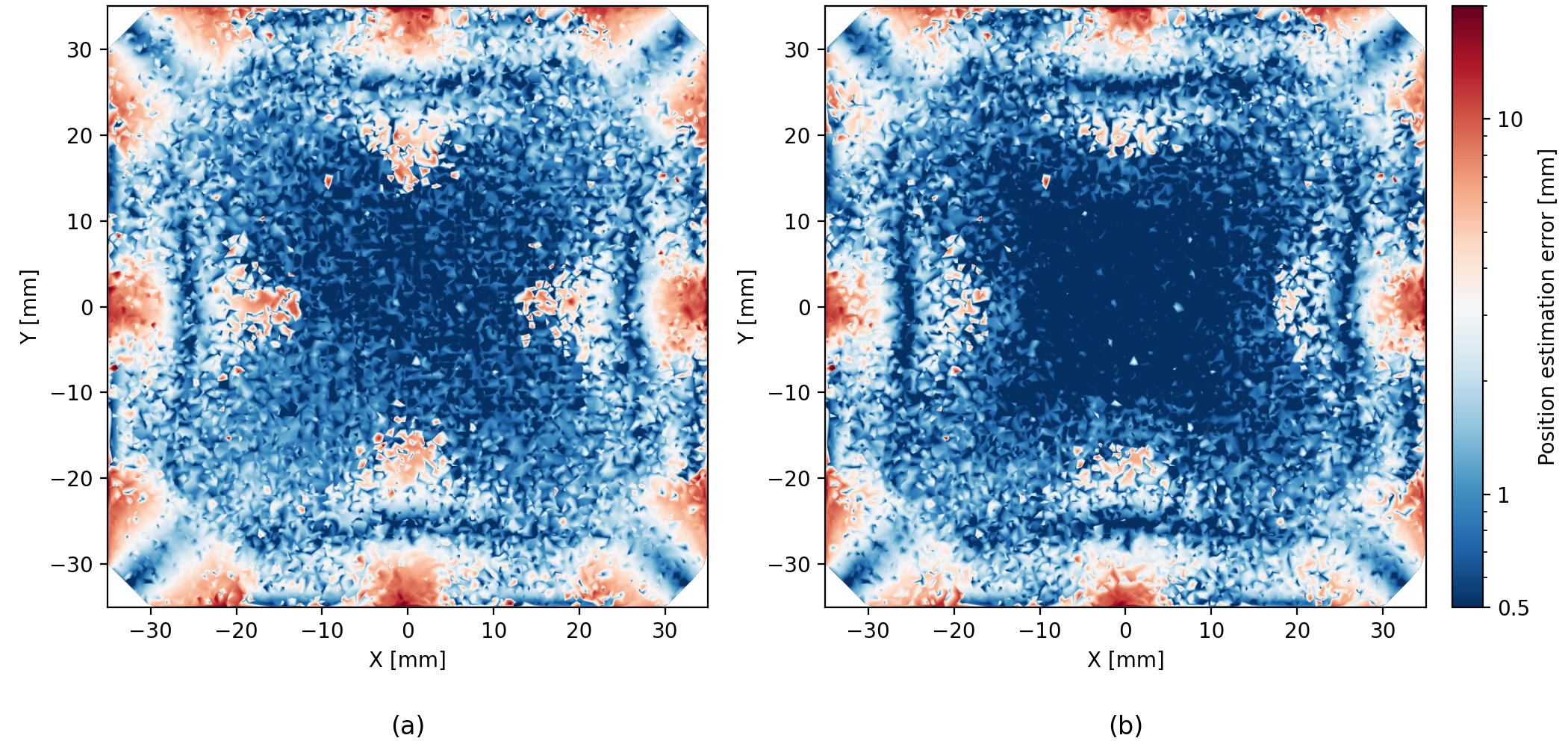}
  \caption{(b) Localization error heatmap for the hybrid probing model assuming a theoretical perfect classifier, i.e., a classifier that always selects the model yielding the most accurate estimate for each measurement. For convenience, subfigure (a) reproduces the localization error map already shown in Figure \ref{fig:oxygen_sim_lgb}.}
    \label{fig:oxygen_sim_lgb_genie}
\end{figure}

To evaluate the upper bound for the hybrid probing method, we determined the mean error achievable if the classifier were flawless. In other words, we computed the error presuming the optimal outcomes of both the ML model and the analytical model predictions were considered. This approach yields a mean error of $1.6$mm. For the localization error heatmap see Figure \ref{fig:oxygen_sim_lgb_genie}.

\subsection{Error distribution}
Error heatmaps reveal that analytic residuals grow near the scintillator's edges and corners, consistent with the GDOP distribution. Machine learning reduces errors globally but is prone to variance. The probing method suppresses edge artifacts by routing to ML, while the boosting hybrid smooths systematic deviations across the entire surface.

%
\section{Discussion and Conclusions}
%
%
We extended the SSPD analytic framework by integrating machine learning and hybrid approaches, achieving a localization accuracy of $\sim 2.0\,$mm (RMSE) and robust LET estimation with $98\%$ accuracy for high-energy oxygen ions impinging on a detector with dimensions of 70~mm$~\times$~70~mm~$\times~$6.7~mm, across a range of incidence positions and angles. For muons we have achieved $\sim 4.0$\,mm RMSE with LET estimation accuracy of $96\%$.
These results suggest that hybrid physics+ML methods offer a strong candidate for future particle detectors.
\\
Each reconstruction strategy offers distinct advantages and limitations.
The analytic model is interpretable, fast, and grounded in photon transport physics, making it ideal for deployment on resource-limited hardware. Its weakness lies in systematic edge biases.
XGBoost offers strong modeling of nonlinearities, but is computationally heavier and less efficient for large-scale data.
LightGBM provides faster training and better scalability, with slightly improved results compared to XGBoost. In our tests, this performance gain was verified empirically, confirming that LightGBM trains significantly faster for the considered datasets.
The hybrid probing approach leverages the strengths of both physics and machine learning: the analytic model is used where it provides accurate predictions, and machine learning fills in the gaps elsewhere. This improves the overall accuracy and reduces the average computational load of inference.

\bibliographystyle{unsrt}
\bibliography{references}

\end{document}